\def\be{\begin{equation}}
\def\ee{\end{equation}}
\def\ba{\begin{array}}
\def\ea{\end{array}}
\def\Nb{{I\!\! N}}
\def\Rb{{I\!\! R}}

\def\Cb{{\Bbb C}}
\documentclass[aps,pra,showpacs,showkeys,preprint]{revtex4}
\usepackage{amsmath}
\usepackage{amssymb}
\usepackage{graphicx}
\begin{document}

\title{Differential Geometry of Bipartite Quantum States}

\author{Zuhuan Yu$^{a}$\footnote{e-mail: yuzh@mail.cnu.edu.cn}}
\author{Xianqing Jost-Li$^{b}$\footnote{e-mail: Xianqing.Li-Jost@mis.mpg.de}}
\author{Qingzhong Li$^{a}$\footnote{e-mail: liqzh@mail.cnu.edu.cn}}
\author{Jintao Lv$^{a}$\footnote{e-mail: lvjintao@sohu.com}}
\author{Shao-Ming Fei$^{a,b}$\footnote{e-mail: fei@wiener.iam.uni-bonn.de}}

\affiliation{$~^{a}$ Department of Mathematics, Capital Normal University, Beijing
100037, China\\
$~^{b}$ Max Planck Institute for Mathematics in the Sciences, D-04103 Leipzig,
Germany}

\begin{abstract}
We investigate the differential geometry of bipartite quantum states.
In particular the manifold structures of pure bipartite states
are studied in detail. The manifolds with respect to
all normalized pure states of arbitrarily given Schmidt ranks
or Schmidt coefficients are explicitly presented.
The dimensions of the related manifolds are calculated.

\end{abstract}

\maketitle

\bigskip
\medskip

Quantum entanglement constitutes the most important resource in quantum information processing
such as quantum teleportation, dense coding, quantum cryptography, quantum error correction and quantum
repeater \cite{nielsen}. The marvelous properties of quantum entanglement are from the
special structures of the multipartite quantum states. Great efforts have been focused on
the proper description and quantification of quantum entanglement \cite{Eof}, the
separability \cite{Sep}, the equivalence of quantum states under local unitary transformations or under
stochastic local operations and classical communication (SLOCC)
for multipartite quantum systems \cite{Eq}.

The geometry of quantum states on a single vector space has been discussed
in \cite{1,2} recently. Let ${\mathcal{H}}$ be an $n$-dimensional complex Hilbert space.
The space of density matrices on ${\mathcal{H}}$, $D({\mathcal{H}})$,
is naturally a manifold stratified
space with the stratification induced by the rank of the state.
The space of all density matrices with rank $r$, $D^{r}({\mathcal{H}})$,
$r=1,2,\cdots,n$, is a smooth and connected manifold of real dimension
$2nr-r^{2}-1$. In particular, $D^{1}({\mathcal{H}})$ is the set of pure states.
Every element of $D({\mathcal{H}})$ is a convex combination of points from
$D^{1}({\mathcal{H}})$. It is shown that $D^{1}({\mathcal{H}})$ is a complex
manifold which is isomorphic to the $n-1$ dimensional complex projective space,
$D^{1}({\mathcal{H}})\simeq CP^{n-1}$, with a metric $g$ determined by the inner product
$\langle M,N\rangle=\frac{1}{2}TrMN$ for density matrices $M$ and $N$.
One can define the Hermitian structure $h$ on $D^{1}({\mathcal{H}})$ by $g$.
In fact, by straightforward calculation, we have
$$
h^{(\alpha)}=\displaystyle \sum_{k,j}h^{(\alpha)}_{kj}{\mathrm{d}}z_{k}
\otimes {\mathrm{d}}\overline{z_{j}}\ ,\ \ h^{(\alpha)}=h_{|D_{\alpha}},~~~\alpha=1,...,n,
$$
where
$$
h^{(\alpha)}_{kj}=
\displaystyle\frac{(1+\displaystyle\sum^{n}_{l=1,l\neq\alpha}|z_{l}|^{2})\delta_{kj}-z_{j}\overline{z_{k}}}
{(1+\displaystyle\sum^{n}_{l=1,l\neq\alpha}|z_{l}|^{2})},
$$
$D_{\alpha}$ is the $\alpha$-th coordinate chart with local complex coordinates $z$ and $\overline{z}$.
Hence it is clear that $h$ differs from the Fubini-Study metric on $CP^{n-1}$ by a constant
multiple.

The quantum entanglement concerns composite systems. In \cite{3} the entanglement
has been discussed in the view of geometry.
In this paper we investigate the manifold structures and classification of pure bipartite states.
We consider quantum states on ${\mathcal{H}}={\mathcal{H}}_{1}\otimes{\mathcal{H}}_{2}$,
where ${\mathcal{H}}_{1}$ and ${\mathcal{H}}_{2}$ are
respectively $n$ and $m$ ($n\leq m$) dimensional complex Hilbert spaces.
We present the explicit manifold constituted by the states with certain Schmidt ranks
or with given Schmidt coefficients, and calculate the dimensions of the related manifolds.

For the convenience, in the following in stead of $|x\rangle$, we simply
denote $x$ as a vector in ${\mathcal{H}}$ and denote $D^{1}({\mathcal{H}})$ as the set of all $x\in{\mathcal{H}}$.
For any $x \in{\mathcal{H}}$, $x$ can be written as the summation of tensor
products,
\be\label{x}
x=x_{1}\otimes y_{1}+x_{2}\otimes
y_{2}+\cdots+x_{k}\otimes y_{k},~~k\in\Nb,
\ee
where $x_{i}\in {\mathcal{H}}_{1}$, $y_{i}\in {\mathcal{H}}_{2}$.
We call the expression (\ref{x}) linearly independent
if $x_{1}$, $x_{2},\cdots,x_{k}$; $y_{1}$, $y_{2},\cdots,y_{k}$ are linearly
independent vectors respectively. We say the length of $x$ is $k$ if (\ref{x}) is
a linearly independent expression.
In fact one can easily prove that the length is just the Schmidt rank
and the Schmidt decomposition is a special expression of a linearly
independent one. Therefore the length of $x$ in all linearly independent expressions
is the same and the terms of tensor products contained in the linearly independent
expression of $x$ are the least in all other possible expressions of $x$.

{\sf [Lemma]} If $x\in {\mathcal{H}}={\mathcal{H}}_{1}\otimes{\mathcal{H}}_{2}$
has the following two linearly independent expressions
\be\label{xx}
x=x_{1}\otimes y_{1}+x_{2}\otimes y_{2}+\cdots+x_{k}\otimes y_{k},~
x=w_{1}\otimes z_{1}+w_{2}\otimes z_{2}+\cdots+w_{k}\otimes z_{k},
\ee
then there exists a non-degenerate $k\times k$ matrix $C$ such that
\be\label{zyc}
(z_{1},\cdots,z_{k})=(y_{1},\cdots,y_{k})C,~~
(w_{1},\cdots,w_{k})=(x_{1},\cdots,x_{k})(C^{t})^{-1}.
\ee

{[\sf Proof]} Expanding $x_{1},x_{2},\cdots,x_{k}$ to be the basis
$x_{1},\cdots,x_{k},x_{k+1},\cdots,x_{n}$ in ${\mathcal{H}}_{1}$ and
$y_{1},y_{2},\cdots,y_{k}$
to be the basis $y_{1},\cdots,y_{k},y_{k+1},\cdots,y_{m}$
in ${\mathcal{H}}_{2}$, we have
$$
w_{j}=\displaystyle\sum^{n}_{i=1}a_{ji}x_{i} \
,z_{j}=\displaystyle\sum^{m}_{i=1}b_{ji}y_{i}
$$
for some $a_{ji}$, $b_{ji}\in\Cb$.
Then from (\ref{xx}) we have
\begin{equation}\label{e:CX}
\displaystyle\sum_{i=1}^{n}\displaystyle\sum_{s=1}^{m}(\displaystyle\sum_{j=1}^{k}
a_{ji}b_{js})x_{i}\otimes
y_{s}=\displaystyle\sum_{j=1}^{k}x_{j}\otimes y_{j}.
\end{equation}

Denote $A$ (resp. $B$) the matrix with entries $a_{ij}$ (resp. $b_{ij}$).
As $\{x_{i}\otimes y_{s}:~j=1,2,\cdots,n;~s=1,2,\cdots,m\}$ is a
basis of ${\mathcal{H}}_{1}\bigotimes {\mathcal{H}}_{2}$, from (\ref{e:CX})
we have
\be\label{atb}
A^{t}B=\left (
\begin{array}{cc}
\displaystyle
E_{k}& 0\\
0&0\\
\end{array}
\right )_{n\times m},
\ee
where $E_{k}$ is the identity matrix of order $k$. If we rewrite $A$ and $B$ as block
matrices $A=(A_{kk}\hspace{3mm} A_{k,n-k})$, $B=(B_{kk}\hspace{3mm} B_{k,m-k})$, then (\ref{atb})
gives rise to $A_{kk}^{t}B_{kk}=E_{k}$, $B_{k,m-k}={0}$, $A_{k,n-k}={0}$. Namely,
$A=(A_{kk}\hspace{4mm}{0)},B=(B_{kk}\hspace{4mm}{0)}$.
Set $C=A^{-1}_{kk}$, we obtain $B_{kk}=(A_{kk}^{t})^{-1}$ and
$
(w_{1},\cdots,w_{k})=(x_{1},\cdots,x_{k})(C^{t})^{-1},
$
$
(z_{1},\cdots,z_{k})=(y_{1},\cdots,y_{k})B^{t}_{kk}=(y_{1},\cdots,y_{k})C.
$
\hfill $\Box$

{\sf [Theorem 1]} Let $D_{k}^{1}({\mathcal{H}})$, a submanifold of
$D^{1}({\mathcal{H}})$, be the set of all normalized pure states with length $k$,
$D_{k}^{1}({\mathcal{H}})=\{x\in {\mathcal{H}},
\mbox{ the length of }x \mbox{ is } k,\|x\|^{2}=1\}$.
We have
$$D_{k}^{1}({\mathcal{H}})\simeq
G(n,k)\times (CP^{k^{2}-1}\displaystyle\backslash
\overline{M})\times G(m,k),
$$
where $\overline{M}$ is a hypersurface of $CP^{k^{2}-1}$, $G(n,k)$ is the Grassmannian manifold.

{\sf [Proof]}
We first prove that there is a one-to-one correspondence between
$D^{1}_{k}({\mathcal{H}})$ and $ G(n,k)\times
(CP^{k^{2}-1}\backslash \overline{M})\times G(m,k)$.

For $x\in D^{1}_{k}({\mathcal{H}}),$ suppose
$x=x_{1}\otimes y_{1}+x_{2}\otimes y_{2}+\cdots+x_{k}\otimes y_{k}$
is a linearly independent expression of $x$. Because
$y_{1},\cdots,y_{k}$ are linearly independent,
$y_{1},\cdots,y_{k}$ span a $k$-dimensional subspace $D_{k}$ of
${\mathcal{H}}_2$. We fix an orthonormal basis
$y_{1}^{0},\cdots,y_{k}^{0}$ in $D_{k}$ and assume
$(y_{1}^{0}\cdots y_{k}^{0})=(y_{1}\cdots y_{k})A,$ where $A$ is a
non-degenerate complex $k\times k$ matrix. If we keep $x$ unchanged,
from Lemma $x_{1},\cdots,x_{k}$ are transformed correspondingly to
$x_{1}^{\prime},\cdots,x_{k}^{\prime},$
$(x_{1}^{\prime}\cdots x_{k}^{\prime})=(x_{1}\cdots
x_{k})(A^{t})^{-1}$.

A $k$-dimensional subspace of
${\mathcal{H}}_{2}$ just corresponds to a point in a Grassmannian
manifold $G(m,k)$. As $x_{1}^{\prime},\cdots,x_{k}^{\prime}$ in the expression
$x=x_{1}^{\prime}\otimes y_{1}^{0}+\cdots+x_{k}^{\prime}\otimes y_{k}^{0}$ are linearly
independent, they span a $k$-dimensional subspace $C_{k}$ of ${\mathcal{H}}_{1}$.
If we fix an orthonormal basis $x_{1}^{0},\cdots,x_{k}^{0}$ in $C_{k},$
then there exists a unique non-degenerate $k\times k$ matrix $G$ such that
$(x_{1},\cdots,x_{k})= (x_{1}^{0},\cdots,x_{k}^{0})G.$
A $k$-dimensional subspace of
${\mathcal{H}}_{1}$ just corresponds to a point in a Grassmannian
manifold $G(n,k)$. Suppose $(x_{1}^{\prime},\cdots,x_{k}^{\prime})=(x_{1}^{0},\cdots,x_{k}^{0})B$,
where $B$ is a $k\times k$ complex matrix with entries $b_{ij}$ satisfying
$\displaystyle\sum^{k}_{i,j=1}|b_{ij}|^{2}=1$, $det(B)\neq0.$ Then all
$B=(b_{ij})_{i,j=1}^{k}$ constitute a set $D$ which can be
viewed as a subset of the identity ball $S^{k^{2}-1}$ in
${\mathbf{C}}^ {k^{2}}$, where
$$S^{k^{2}-1}=\{(b_{11},b_{21},\cdots,b_{k1},b_{12},\cdots,b_{k2},\cdots,b_{kk}):
\displaystyle\sum^{k}_{i,j=1}|b_{ij}|^{2}=1,b_{ij}\in \Cb \}.$$
Moreover, $D$ is an open subset in $S^{k^{2}-1}$.

In summary, to determine $x_{1}^{\prime},\cdots,x_{k}^{\prime}$,
we need to determine the $k$-dimensional subspace $C_{k}$ which is
spanned by $x_{1}^{\prime},\cdots,x_{k}^{\prime}$ and the
nondegenerate $k\times k$ matrix $B$ associated with
$x_{1}^{\prime},\cdots,x_{k}^{\prime}$, i.e. a point of Grassmannian
manifold $G(n,k)$ and a point of $D$ are determined.

We define $A\sim B$ iff there exists $\theta \in \Rb$ such that
$A=e^{i\theta}B$ and denote the equivalence class containing $A$ by
$[A],$ then
$$S^{k^{2}-1}/\sim=CP^{k^{2}-1}.$$
Define $$\begin{array}{cccc} \displaystyle
\pi: &S^{k^{2}-1}&\longrightarrow &CP^{k^{2}-1}\\
&A&\longrightarrow &[A].
\end{array}$$
Then $\pi$ is an open map. Suppose the image of $D$ under $\pi$ is
$\overline{D}$ which is an open subset of $CP^{k^{2}-1}$, so it is
an open submanifold. Suppose  $\overline{M}=CP^{k^{2}-1}\backslash
\overline{D},$ i.e. $\overline{M}$ is the image of the set under the
map $\pi$ which consists of the points contained in $S^{k^{2}-1}$
satisfying $det(B)=0$ and $\overline{M}$ is a hypersurface of
$CP^{k^2-1}$. So we have
$$\overline{D}=CP^{k^{2}-1}\backslash \overline{M}.$$
As
$
(e^{i\theta}x_{1}^{\prime},\cdots,e^{i\theta}x_{k}^{\prime})
=e^{i\theta}(x_{{1}}^0,\cdots,x_{{k}}^0)B=(x_{{1}}^0,\cdots,x_{{k}}^0)(e^{i\theta}B),
$
the action of $e^{i\theta}$ on $x$ can be viewed as on matrix $B$ associated with
$x_{1}^{\prime},\cdots,x_{k}^{\prime}$. So the equivalence class
$[x]$ containing $x$ corresponds to the equivalence class $[B]$
which contains $B$, i.e. $[x]$ corresponds a point in
$\overline{D}$. Hence, a pure state $x$
corresponds to a unique point $p$ in $G(n,k)\times
(CP^{k^{2}-1}\backslash \overline{M})\times  G(m,k)$, where the
coordinates of $p$ are determined uniquely by the $k$-dimensional
subspace $D_{k}$ in ${\mathcal{H}}_{2}$ spanned by
$y_{1},\cdots,y_{k}$, the $k$-dimensional subspace $C_{k}$ in
${\mathcal{H}}_{1}$ spanned by $x_{1},\cdots,x_{k}$ and $[B]$. We
denote this kind of correspondence as $F$. One can
easily prove that $F$ is surjective and injective. So we get a
one-to-one correspondence between $ D^{1}_{k}({\mathcal{H}})$
and $G(n,k)\times (CP^{k^{2}-1}\backslash \overline{M}) \times
G(m,k)$. Moreover from the above proof we know that $F$ is smooth.

We now imbed $G(n,k)\times (CP^{k^{2}-1}\backslash
\overline{M}) \times G(m,k)$ to $CP^{mn-1}$ according to $F$.
For arbitrary $p\in G(n,k)\times (CP^{k^{2}-1}\backslash
\overline{M})\times G(m,k),$ the coordinates of $p$ have the form,
$$
(x_{1,k+1},\cdots,x_{2n},\cdots,x_{kn},a_{12},\cdots,
a_{kk},y_{1,k+1},\cdots,y_{km}).
$$
Let us write the coordinates $(x_{1,k+1},x_{1,k+2},\cdots,x_{kn})$ in $G(n,k)$ in the matrix form
$$
X=\left (
\begin{array}{ccccccc}
\displaystyle
1&0& \cdots &0&x_{1,k+1}&\cdots&x_{1n}\\
0&1& \cdots &0&x_{2,k+1}&\cdots&x_{2n}\\
\cdots&\cdots&\cdots&\cdots&\cdots&\cdots&\cdots\\
0&0& \cdots &1&x_{k,k+1}&\cdots&x_{kn}\\
\end{array}
\right ),
$$
and the coordinates $(a_{12},\cdots, a_{1k},\cdots ,a_{kk})$ in $CP^{k^{2}-1}$
in the form
$$A=\left (
\begin{array}{cccc}
\displaystyle
1&a_{12}& \cdots &a_{1k}\\
a_{21}&a_{22}&\cdots &a_{2k}\\
\cdots&\cdots&\cdots&\cdots\\
a_{k1}&a_{k2}& \cdots &a_{kk}\\
\end{array}
\right ),$$

Set $X^{t}A=B$. Then $B=(b_{ij})$ is an $n\times k$ matrix.
Let $e_{1},\cdots,e_{n}$ (resp. $d_{1},\cdots,d_{m}$) be an orthonormal basis in
${\mathcal{H}}_{1}$ (resp. ${\mathcal{H}}_{2}$). Take
$$
x_{1}=\displaystyle\sum^{n}_{j=1}b_{j1}e_{j},~x_{2}=\displaystyle\sum^{n}_{j=1}b_{j2}e_{j},~\cdots,~
x_{k}=\displaystyle\sum^{n}_{j=1}b_{jk}e_{j},
$$
and
$$
y_{1}=\displaystyle\sum^{m}_{j=1}y_{1j}d_{j},~y_{2}=\displaystyle\sum^{m}_{j=1}y_{2j}d_{j},~\cdots,~
y_{k}=\displaystyle\sum^{m}_{j=1}y_{kj}d_{j},
$$
where $y_{ij}$ are the entries of the matrix $Y$,
$$Y=\left (
\begin{array}{ccccccc}
\displaystyle
1&0& \cdots &0&y_{1,k+1}&\cdots&y_{1m}\\
0&1& \cdots &0&y_{2,k+1}&\cdots&y_{2m}\\
\cdots&\cdots&\cdots&\cdots&\cdots&\cdots&\cdots\\
0&0& \cdots &1&y_{k,k+1}&\cdots&y_{km}\\
\end{array}
\right ),
$$
then both $x_{1},\cdots,x_{k}$ and $y_{1},\cdots,y_{k}$ are linearly independent respectively.

Let
$$
x=x_{1}\otimes y_{1}+\cdots+x_{k}\otimes y_{k}
=\displaystyle \sum^{n}_{j=1}\displaystyle
\sum^{m}_{s=1}(\displaystyle \sum^{k}_{l=1}b_{jl}y_{ls})
e_{j}\otimes d_{s}.
$$
Then $x\in D^{1}_{k}({\mathcal{H}})$ is the image of $p$ under $F$. Since
${\mathcal{H}}_{1}\simeq\Cb^{n}$, ${\mathcal{H}}_{2}\simeq \Cb^{m}$ and
$D^{1}({\mathcal{H}})\simeq CP^{nm-1},$ we can define the imbedding:
$$\begin{array}{cccc}
f:&G(n,k)\times (CP^{k^{2}-1}\backslash \overline{M})\times G(m,k)&\longrightarrow&CP^{mn-1}\\
&p&\longrightarrow&q
\end{array}$$
where $q=f(p)=x$. The homogeneous coordinates of
$q$ are given by  $q=(d_{11},d_{12},\cdots,d_{1m},d_{21},\cdots,d_{2m},\cdots,d_{nm}),$
where $d_{js}=\displaystyle \sum^{k}_{l=1}b_{jl}y_{ls}=
\displaystyle \sum^{k}_{l=1}\displaystyle
\sum^{k}_{t=1}x_{tj}a_{tl}y_{ls}$ $(j=1,\cdots,n,s=1,\cdots,m).$ Then
the coordinate components of $q$ are polynomial of the coordinate
components of $p$. Hence, the imbedding $f$ is non-degenerate
holomorphic mapping. Moreover, we have
$$
f(G(n,k)\times
(CP^{k^{2}-1}\backslash \overline{M})\times G(m,k))=
D^{1}_{k}({\mathcal{H}}).
$$
Therefore $D^{1}_{k}({\mathcal{H}})$ is a
complex submanifold of $CP^{mn-1}$ ( i.e. $D^{1}({\mathcal{H}})$), and
$$D^{1}_{k}({\mathcal{H}})\simeq G(n,k)\times (CP^{k^{2}-1}\backslash \overline{M})\times G(m,k).$$
\hfill $\Box$

{\sf Theorem 2} The subset
$D^{1}_{k}(\mu_1,\cdots ,\mu_k)$ of $D_{k}^{1}({\mathcal{H}})$ of
pure states with the Schmidt coefficients
$\mu_1 \geqslant \mu_2 \geqslant \cdots \geqslant \mu_k$
is a submanifold of real dimension $2k(m+n-k)-k-1$, which is diffeomorphically
equivalent to a manifold
$$(CP^{n-1}\times CP^{m-1})\times \cdots \times (CP^{n-k}\times
CP^{m-k})\times T^{k-1},$$ where $T^{k-1}$ is a torus of real
dimension $k-1$.

{\sf [proof]}
For any pure state $[e]$ of $D^{1}_{k}({\mathcal{H}})$, the unit vector $e$ has the
following Schmidt representation
$e=\mu_1 a_1\otimes b_1+\cdots+\mu_k a_k\otimes b_k,$
where $a_i^{,}s$ and $b_i^{,}s$ are orthonormal vectors in
$\mathcal{H}_1$ and $\mathcal{H}_2$ respectively, and $\mu_i^{,}s$
are Schmidt coefficients of $e$, we assume that
$\mu_1\geqslant \mu_2 \geqslant \cdots \geqslant \mu_k.$
Consider the element $\tilde{e}$ which has the same Schmidt
coefficients as $e$,
$\tilde{e}=\mu_1 \tilde{a_1}\otimes \tilde{b_1}+\cdots+\mu_k \tilde{a_k}\otimes \tilde{b_k},$
and $[a_i]=[\tilde{a_i}]\in D^{1}(\mathcal{H}_{1}),\ \
[b_i]=[\tilde{b_i}]\in D^{1}(\mathcal{H}_{2}),\ \ i=1,\cdots,k.$ Hence
$\tilde{e}$ must have the form
$\tilde{e}=\mu_1 e^{i\theta_{1}}a_1 \otimes b_1+\cdots+\mu_k e^{i\theta_{k}}a_k\otimes b_k,$
and $[\tilde{e}]$ constitute a set
$$\{([a_1],[b_1],\cdots,[a_k],[b_k],e^{i\beta_{1}},\cdots,e^{i\beta_{k-1}})\ |\ \beta_{1},
\cdots,\beta_{k-1}\in \Rb \}\simeq T^{k-1}.
$$
Then all the pure states with the same Schmidt coefficients
$\mu_1\geqslant \mu_2\geqslant\cdots\geqslant\mu_k$
constitute a set which is equivalent to a manifold
$(CP^{n-1}\times CP^{m-1})\times \cdots \times (CP^{n-k}\times
CP^{m-k})\times T^{k-1},$ which is of real dimension
$2k(m+n-k)-k-1$.
\hfill $\Box$

As a simple example, let us first take
$dim({\mathcal{H}}_{1})=dim({\mathcal{H}}_{2})=3$, $k=1$.
For arbitrary $x\in D^{1}({\mathcal{H}}_{1}),$ $y\in
D^{1}({\mathcal{H}}_{2})$, by the Segre imbedding we have
$Seg(x,y)=|x\otimes y\rangle\langle x\otimes y|.$
As $w=x\otimes y\in D^{1}_{1}({\mathcal{H}})$, one gets
$$
Seg(D^{1}({\mathcal{H}}_{1})\times
D^{1}({\mathcal{H}}_{2}))\subset D^{1}_{1}({\mathcal{H}}).
$$
And for arbitrary $w\in D^{1}_{1}({\mathcal{H}})$,
there exist $x\in {\mathcal{H}}_{1}$, $y\in {\mathcal{H}}_{2}$ such
that $w=x\otimes y=Seg(x,y).$
The Segre imbedding
$Seg:D^{1}({\mathcal{H}}_{1})\times D^{1}({\mathcal{H}}_{2})\rightarrow D^{1}_{1}({\mathcal{H}})$
is a surjective map to $D^{1}_{1}({\mathcal{H}})$. Hence, we have
$Seg(D^{1}({\mathcal{H}}_{1})\times D^{1}({\mathcal{H}}_{2}))=D^{1}_{1}({\mathcal{H}})$.
Therefore $D^{1}_{1}({\mathcal{H}})\cong D^{1}({\mathcal{H}}_{1})\times
D^{1}({\mathcal{H}}_{2})\simeq CP^{2}\times CP^{2}.$ From
Theorem 1, in this case $k^{2}-1=0$. We get
$D^{1}_{1}({\mathcal{H}})\simeq CP^{2}\times CP^{2}$.

As a more complicated case, we consider
$dim({\mathcal{H}}_{1})=3$, $dim({\mathcal{H}}_{2})=4,$ and $k=2.$
For arbitrary $p\in G(3,2)\times (CP^{3}\backslash
\overline{M})\times G(4,2)$ with coordinate
$p=(x_{13},x_{23},$ $a_{12},a_{21},$ $a_{22},y_{13},$ $y_{14},y_{23},y_{24})$, set
$$
X=\left (\begin{array}{ccc}
1&0&\displaystyle x_{13}\\
0&1&x_{23}
\end{array}
\right ),~~
A=\left(
\begin{array}{cc}
1&a_{12}\\
a_{21}&a_{22}
\end{array}
\right ),~~
Y= \left(
\begin{array}{cccc}
1&0&y_{13}&y_{14}\\
0&1&y_{23}&y_{24}
\end{array}
\right ).
$$
Then
$$X^{t}A=\left(
\begin{array}{cc}
1&a_{12}\\
a_{21}&a_{22}\\
x_{13}+a_{21}x_{23}&a_{12}x_{13}+x_{23}a_{22}
\end{array}
\right ).$$
We take
$x_{1}=e_{1}+a_{21}e_{2}+(x_{13}+a_{21}x_{23})e_{3},
x_{2}=a_{12}e_{1}+a_{22}e_{2}+(a_{12}x_{13}+x_{23}a_{22})e_{3},$
$y_{1}=d_{1}+y_{13}d_{3}+y_{14}d_{4},y_{2}=d_{2}+y_{23}d_{3}+y_{24}d_{4}.$
Let $x=x_{1}\otimes y_{1}+x_{2}\otimes y_{2}.$ Since
$x_{1},x_{2}$; $y_{1},y_{2}$ are linearly independent respectively, we
have  $x\in D_{2}^{1}(\mathcal{H}).$

For arbitrary $x\in D_{2}^{1}(\mathcal{H})$,
suppose $x=x_{1}\otimes y_{1}+x_{2}\otimes y_{2},$ then
$y_{1}$, $y_{2}$ span a unique 2-dimensional subspace $D_{2}$ of
${\mathcal{H}}_2$. We fix an orthonormal basis $y_{1}^{0}$, $y_{2}^{0}$
in $D_{2}$ and suppose $(y_{1}^{0}, y_{2}^{0})=(y_{1},y_{2})A,$
where $A$ is a non-degenerate complex $2\times 2$ matrix. At the same
time, suppose that $x_{1},~x_{2}$ are transformed correspondingly to
$x_{1}^{\prime},~x_{2}^{\prime}$, $(x_{1}^{\prime},
x_{2}^{\prime})=(x_{1},x_{2})(A^{t})^{-1}$. Then
$x=x_{1}^{\prime}\otimes y_{1}^{0}+x_{2}^{\prime}\otimes y_{2}^{0},$
and $x_{1}^{\prime},~x_{2}^{\prime}$ generate a unique 2-dimensional
subspace $C_{2}$ of ${\mathcal{H}}_1$. We fix an orthonormal basis
$x_{1}^{0},x_{2}^{0}$ in $C_{2}$  and assume $(x_{1}^{\prime},
x_{2}^{\prime})=(x_{1}^{0},x_{2}^{0})B$. Then
$(x_{1}^{\prime},x_{2}^{\prime})$ are determined uniquely by $C_{2}$
and $B$. Moreover, $[x_{1}^{\prime},x_{2}^{\prime}]$ correspond to
$[B]$, and $[B]\in CP^{3}\backslash \overline{M},$ where
$\overline{M}=\{[A]:~A\mbox{ are complex $2\times 2$ matrices},
det(A)=0\}$. $C_{2}$ is associated to a point of Grassmannian
manifold $G(3,2)$ and $D_{2}$ is associated to a point of
Grassmannian manifold $G(4,2),$ i.e. $x$ is associated to a point of $G(3,2)\times
 (CP^{3}\backslash \overline{M})\times G(4,2)$.

Furthermore, for arbitrary point in $G(3,2)\times
( CP^{3}\backslash \overline{M})\times G(4,2)$, we can find
correspondingly a unique point in $D^{1}_{2}(\mathcal{H}),$ and vice versa.
In this case, the imbedding is
$$\begin{array}{cccc}
f:&G(3,2)\times (CP^{3}\backslash \overline{M})\times G(4,2)&\longrightarrow&CP^{11}\\
&p&\longrightarrow&q
\end{array}$$
where $q=f(p)=x$ and the homogeneous coordinates of $q$ are assumed to be
$q=(d_{11},d_{12},d_{13},d_{14},d_{21},d_{22},d_{23},d_{24},d_{31},d_{32},d_{33},d_{34}),$
where
$d_{11}=1,d_{12}=a_{12},d_{13}=a_{12}y_{23}+y_{13},d_{14}=a_{12}y_{24}+y_{14},
d_{21}=a_{21},d_{22}=a_{22},d_{23}=a_{21}y_{13}+a_{22}y_{23},d_{24}=a_{21}y_{14}+a_{22}y_{24},
d_{31}=x_{13}+a_{21}x_{23},d_{32}=a_{12}x_{13}+a_{22}x_{23},d_{33}=y_{23}(a_{12}x_{13}+a_{22}x_{23})
+y_{13}(x_{13}+a_{21}x_{23}),d_{34}=y_{24}(a_{12}x_{13}+a_{22}x_{23})
+y_{14}(x_{13}+a_{21}x_{23}).$

The first example tests the theorem from the
Segre imbedding point of view. In this case the
second factor of the product manifold generates a point. The
second one is a lower dimension case according to the Theorem 1.

We have investigated the complex manifold structure for bipartite pure states and the
K\"ahler metric of $D^{1}_{k}({\mathcal{H}})$, by
presenting explicitly the manifolds with respect to
all pure states of arbitrarily given Schmidt ranks
or Schmidt coefficients and calculating
the dimensions of the corresponding manifolds.
In fact, we also can express the K\"ahler metric of $D^{1}_{k}({\mathcal{H}})$ by local
coordinates, but the expressions are very complicated and it is
difficult to compute the geometrical objects such as holomorphic
curvature, scalar curvature. It would be also nice to describe the entanglement
of quantum states according to some functions of metric or geometrical
objects. The results in this paper can be used to study the differential geometry
of bipartite mixed states.

{\sf Acknowledgment}
We would like to express gratitude to K. Wu for many helpful suggestions and
X.C. Rong, Z.F. Yang for valuable comments.
The work is partly supported by NSFC projects
10371014, 10675086, Funds of China Scholarship Council,
Funds of Beijing YouXiuRenCai and KM200510028022, NKBRPC(2004CB318000).


\begin{thebibliography}{99}

\bibitem{nielsen} M.A. Nielsen and I.L. Chuang, Quantum Computation and
Quantum Information, Cambridge University Press, Cambridge, 2000.

\bibitem{Eof} C.H. Bennett, D.P. DiVincenzo, J.A. Smolin, and W.K.
Wootters, Phys. Rev. A \textbf{54}, 3824 (1996).\\
M. Horodecki, Quant. Inf. Comp. \textbf{1}, 3 (2001).\\
D. Bru\ss , J. Math. Phys. \textbf{43}, 4237 (2002).\\
W.K. Wootters, Phys. Rev. Lett. \textbf{80}, 2245(1998).\\
B.M. Terhal and K.G.H. Vollbrecht, Phys. Rev. Lett. \textbf{85}, 2625 (2000).\\
S.M. Fei and X.Q. Li-Jost, Rep. Math. Phys. \textbf{53}, 195 (2004).\\
K. Chen, S. Albeverio, and S.M. Fei, Phys. Rev. Lett. \textbf{95}, 040504 (2005).\\
K. Chen, S. Albeverio, and S.M. Fei, Phys. Rev. Lett. \textbf{95}, 210501 (2005).
S.M. Fei, Z.X. Wang and H. Zhao, Phys. Lett. {\bf A 329}(2004)414-419.

\bibitem{Sep}
M. Lewenstein, D. Bru{\ss}, J. I. Cirac, B. Kraus, M. Ku\'s, J.
Samsonowicz, A. Sanpera,  and R. Tarrach, J. Mod. Phys. {\bf 47},
2481 (2000).\\
R. Werner, Phys. Rev. A{\bf 40}, 4277 (1989).\\
A. Peres  Phys. Rev. Lett. {\bf 77}, 1413 (1996).\\
M. Horodecki, P. Horodecki, and R. Horodecki, Phys. Lett. A {\bf 223}, 8 (1996).\\
W. D\"ur, G. Vidal and J.I. Cirac, Phys. Rev. A {\bf 62},
062314(2000).\\
S. Karnas and  M. Lewenstein, Phys. Rev. A 64, 042313 (2001).\\
S. Albeverio, S.M. Fei and D. Goswami, Phys. Lett. {\bf A 286} (2001)91-96.\\
S.M. Fei, X.H. Gao, X.H. Wang, Z.X. Wang and K. Wu, Phys. Lett. A {\bf 300} (2002)559-566;
Phys. Rev. {\bf A 68} (2003) 022315.\\
S. Albeverio, K. Chen and S.M. Fei, Phys. Rev. {\bf A 68}(2003)062313.

\bibitem{Eq}
E.M.~Rains, {\em IEEE Transactions on Information Theory} {\bf 46} 54-59(2000).\\
M.~Grassl, M.~R\"otteler and T.~Beth, {\em Phys. Rev. A} {\bf 58}, 1833(1998).\\
Y. Makhlin, {\em Quant. Info. Proc.} {\bf 1}, 243-252 (2002).\\
N. Linden, S. Popescu and A. Sudbery, {\em Phys. Rev. Lett.} {\bf
83}, 243 (1999).\\
S. Albeverio, S.M. Fei, P. Parashar and W.L. Yang, Phys. Rev. A {\bf 68} (Rapid Comm.) (2003) 010303.\\
S. Albeverio, L. Cattaneo, S.M. Fei and X.H. Wang, Int. J. Quant. Inform. {\bf 3}(2005)603-609;
Rep. Math. Phys. {\bf 56} (2005)341-350.\\
S.M. Fei and N.H. Jing, Phys. Lett. {\bf A 342}(2005)77-81.\\
C. H. Bennett, S. Popescu, D. Rohrlich, J. A. Smolin, A. V. Thapliyal, Phys. Rev. A
63 (2001) 012307.\\
W. D\"ur, G. Vidal, J. I. Cirac, Phys. Rev. A 62, 062314(2000).\\
F. Verstraete, J.Dehaene, B.De Moor and H. Verschelde Phys. Rev. A. 65, 052112 (2002).

\bibitem{1}
Janusz Grabowski, Giuseppe Marmo, Maret Kus, Geometry of quantum systems:density states and
entanglement, J. phys. A 38(2005)10217-10244.

\bibitem{2}
V. I. Man'ko, G. Marmo, E. C. G. Sudarshan, F. Zaccaria, Differential geometry
of density states, Rept. Math. Phys. 55(2005)405-422.

\bibitem{3}
J. Grabowski, M. Ku\'s and G. Marmo, Open Sys. and Information Dyn. 13(2006)343-362.

\end{thebibliography}
\end{document}